\documentclass[journal]{IEEEtran}

\usepackage{multirow}
\usepackage{cite}
\usepackage[pdftex]{graphicx}
\usepackage{graphicx}
\usepackage{amsmath}
\usepackage{amssymb}
\usepackage{array}
\usepackage{fixltx2e}

\usepackage{url}
\usepackage{epstopdf}
\usepackage[ruled]{algorithm2e}
\usepackage{color}
\usepackage{comment}
\usepackage{subfigure}
\usepackage{epsfig}

\usepackage[utf8]{inputenc}
\usepackage[T1]{fontenc}

\begin{document}

\title{Artificial Intelligence for 6G Networks: Technology Advancement and Standardization}

\author{Muhammad K. Shehzad, Luca Rose, M. Majid Butt, István Z. Kovács, Mohamad Assaad, and Mohsen Guizani \thanks{Muhammad K. Shehzad (corresponding author) is with Nokia Bell-Labs, France, and Laboratoire des Signaux et Systèmes, CentraleSupelec, CNRS, University of Paris-Saclay, France. (\{muhammad.shehzad\}@nokia.com/centralesupelec.fr).

Luca Rose, and M. Majid Butt are with Nokia Bell-Labs, France. (\{luca.rose, majid.butt\}@nokia-bell-labs.com).

István Z. Kovács is with Nokia Standards, Aalborg, Denmark. (istvan.kovacs@nokia.com).

Mohamad Assaad is with Laboratoire des Signaux et Systèmes, CentraleSupelec, CNRS, University of Paris-Saclay, France. (mohamad.assaad@centralesupelec.fr).

Mohsen Guizani is with Mohamed Bin Zayed University of Artificial Intelligence (MBZUAI), Abu Dhabi, UAE. (mguizani@ieee.org).
}
}

\maketitle

\begin{abstract}
With the deployment of 5G networks, standards organizations have started working on the design phase for sixth-generation (6G) networks. 6G networks will be immensely complex, requiring more deployment time, cost and management efforts. On the other hand, mobile network operators demand these networks to be intelligent, self-organizing, and cost-effective to reduce operating expenses (OPEX). Machine learning (ML), a branch of artificial intelligence (AI), is the answer to many of these challenges providing pragmatic solutions, which can entirely change the future of wireless network technologies. By using some case study examples, we briefly examine the most compelling problems, particularly at the physical (PHY) and link layers in cellular networks where ML can bring significant gains. We also review standardization activities in relation to the use of ML in wireless networks and future timeline on readiness of standardization bodies to adapt to these changes. Finally, we highlight major issues in ML use in the wireless technology, and provide potential directions to mitigate some of them in 6G wireless networks. 
\end{abstract}

\begin{IEEEkeywords}
AI, ML, Wireless networks, 3GPP, 6G.
\end{IEEEkeywords}

\IEEEpeerreviewmaketitle

\section{Introduction}
\begin{figure} 
\begin{center}
  \includegraphics[scale=0.59]{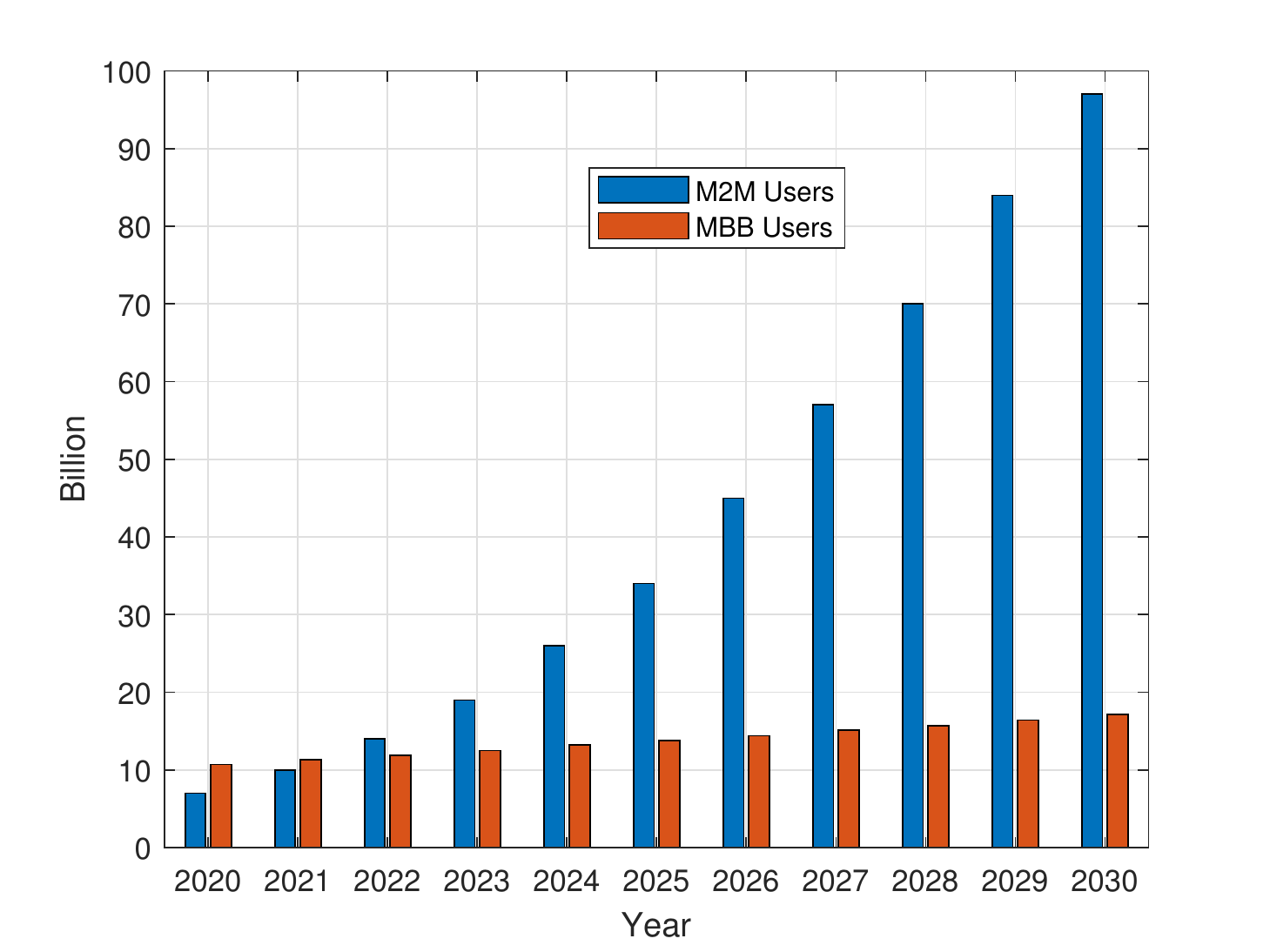}
  \caption{Estimation of global mobile subscriptions in machine-to-machine (M2M) and mobile broadband (MBB) from 2020 to 2030. Source: ITU-R Report M.2370-0 \cite{imt}.}\label{traffic2}.      
\end{center}
\end{figure}

Unprecedented growth in the global cellular traffic (as shown in Fig.\,\ref{traffic2}) and immense data rate demands have become a challenge, leading wireless industry to the next-generation, called 6G. 6G-era will bring digital, physical and biological worlds together with the goal to improve human experience and well-being. 6G will be operating in TeraHertz (THz) frequencies ($0.1$-$10$\,THz), hence beneficial for multiple use cases in industrial applications, providing immense data rates ($\approx1\,$Tb/s), accelerating internet-of-things, and wider network coverage. AI/ML will pave the way for THz communications at different layers \cite{THz_ML}, e.g., supporting channel acquisition \cite{RNN} and modulation classification \cite{Jakob} at PHY. Similarly, at the link layer, beamforming design and channel allocation can exploit ML \cite{THz_ML}. In THz systems, a channel can significantly vary at a micrometer scale, resulting in a tremendous increase in channel estimation frequency and corresponding overhead. ML algorithms can counter this issue by using, e.g., improved channel prediction techniques \cite{RNN, CP_Karam}.

Recently, fast-growing deployment of 5G has opened up many challenges, including massive complexity in network architecture, low latency, high cost, power consumption, and deployment of hybrid Long-Term Evolution (LTE) new radio (NR), leading to difficulties in network optimization. In such a complex scenario, the network intelligence has become a major focus as it will play a pivotal role in complex problem solving \cite{6G_UAV_IoT_AI}, e.g., self-healing, self-optimization, and self-configuration of a network \cite{3GPPAI}.

Future networks will become ``cognitive'' in a way that many aspects such as spectrum sensing/sharing, slicing, radio resource management (RRM), and mobility management, will be ML-based. Further, it is expected that ML will impact 6G air interface fundamentally and it will be designed to support ML natively \cite{Jakob_AI_air}. Several recent research attempts, e.g.,
\cite{6G}, propose different road maps for 6G, but they do not address standardization timeline and related issues regarding application of ML in 6G. Albeit, to some extent, \cite{B5G_AI} gives an overview of ML and standardization; nevertheless, ML-related technical challenges and its applications from an industrial and standardization perspective are not addressed.

\begin{figure*} 
\begin{center}
  \includegraphics[width=18cm,height=18cm]{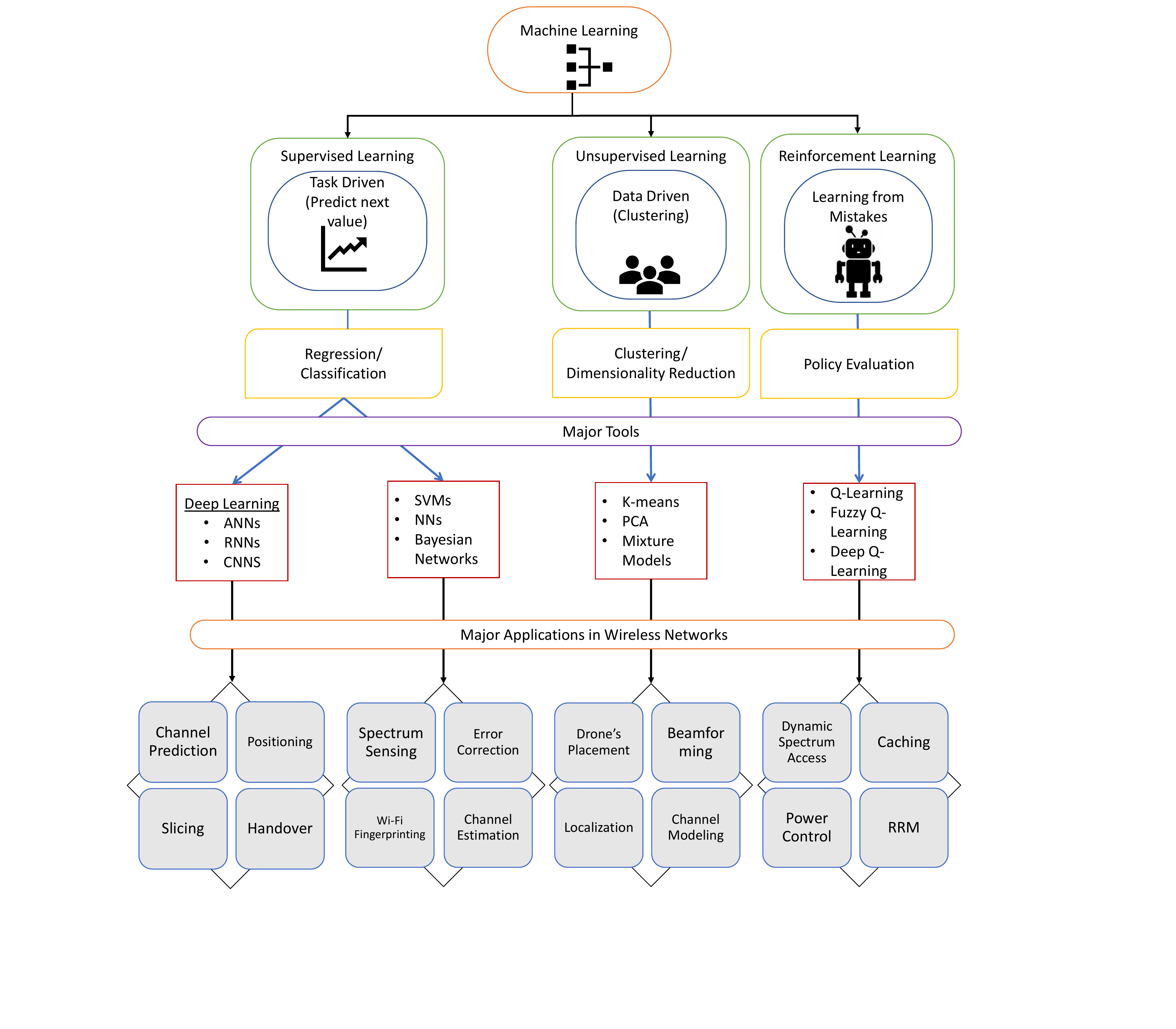}
  \caption{An overview of ML paradigms, major tools, and applications in wireless networks.}\label{Fig2}      
\end{center}
\end{figure*}

Reconfigurable intelligent surface (RIS) and non-orthogonal multiple access (NOMA) are two key technologies for 6G \cite{AI_RIS_NOMA}. RIS can re-engineer electromagnetic waves, hence beneficial to deliver the information where obstacles block the destination. RIS can be integrated with ML, allowing RIS to acquire environmental information by configuring various sensors, while ML can learn dynamic parameters intelligently, reducing the computation cost of RIS-based networks. Similarly, NOMA is a promising access technique for 6G. In ML-empowered NOMA-based networks, gNodeBs (gNB) can intelligently define their control policy and improve decision-making ability.

Today's networks use model-based methods to optimize various network functions providing characteristics of the process involved. However, these models might be too complex to be implemented in a realistic time frame or they include a great level of abstraction to function in a general environment. In contrast, ML-based solutions can adapt to real-time (RT) scenario changes and localized characteristics, learning the specific environment around the transceivers. 
The contributions of this article are twofold:
\begin{itemize}
    \item We look at the above-mentioned problems from an industrial perspective and outline the gap between research and practice.
    \item We review standardization activities in the context of adopting ML in various aspects of wireless communications, e.g., channel acquisition, positioning. Furthermore, we highlight major issues and possible research directions in relation to the use of ML in wireless networks.
\end{itemize}

\section{Overview of ML Techniques in Wireless Networks} \label{sec2}
 
ML is a process of training machines through data without explicit programming. Broadly speaking, ML consists of three paradigms: unsupervised learning, supervised learning, and reinforcement learning (RL). All these paradigms have a \textit{training/exploration phase} to optimize a learning algorithm that later can be used in \textit{prediction/exploitation phase} to infer on unknown inputs. As shown in Fig.~\ref{Fig2}, we briefly summarize them by providing some use cases in wireless networks. 
\subsubsection{Supervised Learning} 
Supervised learning exploits a labelled data set to learn a (hidden) function  that maps an input to an expected output based on the examples. The standard techniques used to solve supervised learning-based problems are artificial neural networks (ANNs), support vector machines (SVMs), Bayesian networks, recurrent neural networks (RNNs), and convolutional neural networks (CNNs).

\subsubsection{Unsupervised Learning} Unsupervised learning does not learn from labelled data, instead, training is based on an unlabelled data set. K-means and principal component analysis (PCA) are examples of two major tools  used for clustering and dimensionality reduction, respectively.

\subsubsection{Reinforcement Learning} RL is not based on training but rather the agent/decision-maker learns and decides online, maximizing a long-term reward. RL is beneficial in control problems where the agent adapts to changing environmental conditions, e.g., uplink power control.

Motivated by the considerable benefits of ML in various fields, its applications have also been considered in wireless networks almost at all layers of communication. Here, we focus on its impact on radio access networks (RAN), particularly PHY and link layers. Based on ML tools, given in Fig.\,\ref{Fig2}, some case studies will be explained later in Section\,\ref{sec3}.

\subsection{{Machine Learning at PHY}} \label{MLPHY}
At PHY, many optimization problems are non-convex, e.g., sum-rate maximization. 
ML is a powerful tool to find good solution(s) for such non-convex optimization problems. Based on advanced learning algorithms, 6G networks provide the following major advantages by using ML.

\begin{itemize}
\item ML can be effective to deal with network complexity. 6G networks will be more complex due to numerous network topologies, immense growth in the cellular users, staggering data rate demands, complex air interface, vast network coordination methods, etc. Forecasting considerable complexity of 6G networks, the derivation of optimum performance solutions is nearly infeasible without ML.
 
\item ML can play a vital role to deal with model deficit problems. Current cellular networks are amenable for mathematical derivation, for instance, information theory gives closed-form expressions for various problems such as Shannon theorem. However, the inherent complexity of 6G networks hinders the possibility of exploiting closed-form analytical expression(s), which can be due, for instance, to non-linearities either in the channel 
or network devices. ML offers an efficient way to deal with non-linearities, providing feasible solution(s) in a tractable manner.

\item ML can cope with algorithm deficit problems. In current cellular networks, many optimal algorithms, although well-characterized, are impractical to be implemented. 
Considering the example of multiple-input multiple-output (MIMO) systems where optimal solutions are known (e.g., dirty paper coding), they are overlooked in favour of linear solutions, e.g., linear minimum mean-squared error. It is envisaged that ML can pave the way to implement more efficient yet practical solutions.

\end{itemize}

ML has been used to study various PHY issues, and without being exhaustive, some of the recent areas include:
\begin{itemize}

    \item CNNs are used for modulation classification in \cite{Jakob}.
    \item An RNN-based wireless channel predictor \cite{CP_Karam} is used in \cite{RNN}, explained in Section\,\ref{CSI_Feedback}, to deal with inaccurate channel state information (CSI).
\end{itemize}

\section{Wireless Networks: Case Studies}\label{sec3}
{In this section, we present three use cases to demonstrate the use of ML techniques in industrial wireless networks. ML tools utilized for these use cases are depicted in Fig.\,2.} 
\subsection{UE Positioning}
Highly accurate user equipment (UE) positioning is one of the prime considerations for Third Generation Partnership Project (3GPP) studies beyond Release\,$15$. Various angle and time-of-arrival-based methods are used to determine UE positioning in today's cellular networks. All of these methods require triangulation techniques to resolve UE position and suffer from time synchronization errors.

\begin{figure}
\centering
\includegraphics[scale=0.66]{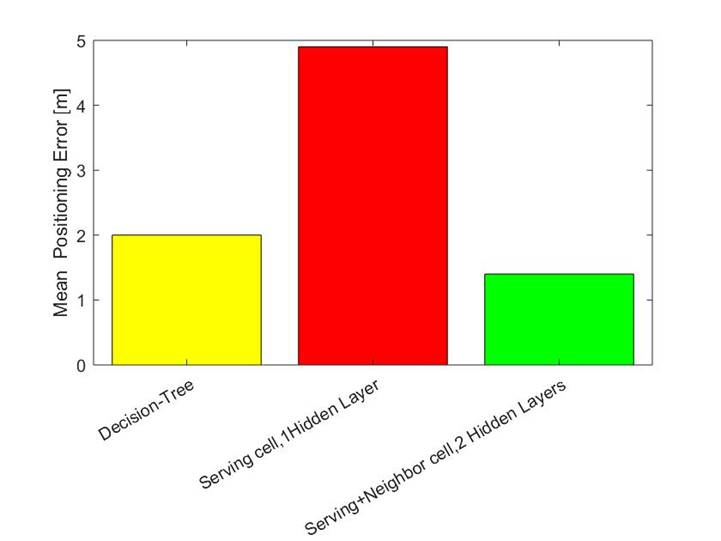}
\caption{Comparison of UE position for both DNN and decision tree techniques. The system level parameters for the network includes $8$ sites with Inter-site distance $110$\,m and carrier frequency $28$\,GHz. For details of the parameters, please refer to \cite{Majid_result}. }
\label{Positionining}
\end{figure}

We studied UE position by using radio frequency (RF) fingerprinting and two ML techniques, namely deep learning and decision tree,  for an outdoor scenario \cite{Majid_result}. Serving cell Reference Signal Received Power (RSRP) as well as neighbor cell RSRP values were used as features to train a deep neural network (DNN). As shown in Fig.\,\ref{Positionining}, nearly $5$\,m accuracy is achieved for DNN when only $4$ serving cell RSRP values and corresponding beam IDs are considered as a feature input, while it improves to nearly $1$\,m when $2$ more RSRP values from the strongest neighboring cells, respective cell and beam IDs are added to the input feature set. The decision tree, a less complex algorithm as compared to DNN, provides about $2$\,m accuracy using data from both serving and neighboring cell beams as an input feature. The mean accuracy of nearly $1$\,m obtained from DNN is comparable to the accuracy level achieved with traditional methods without requiring triangulation and does not suffer from signal timing synchronization issues.

\subsection{ML-Assisted Proactive Mobility}
\label{sec:mobility}

For seamless and efficient mobility, a well optimized network should reduce the number of Handover (HO) events while avoiding Handover Failures (HOF) and Radio Link Failures (RLF). 
An emerging approach is to utilize ML-based algorithms, which enable \textit{proactive} and \textit{UE specific} mobility actions in the gNB. A relatively simple approach to this is to design an ML-based estimator of the radio measurements, such as RSRP of serving and neighbor cells, with a certain minimum accuracy and within a certain time horizon. Radio measurements are traditionally performed at the UEs side and reported to the serving gNB (or gNB-Centralized Unit) according to specific Radio Resource Control (RRC) configurations. For ML-based prediction purposes, time-traces of RSRP, or Reference Signal Received Quality (RSRQ) values need to be collected either in the UE and/or serving the gNB.

For example, collected time-series of RSRP values are used as input to the ML-based predictor, which provides at the UE, and/or at the serving gNB, a set of sufficiently accurately estimated RSRP values within a given future time horizon. Then, these signal estimations are used for predictive evaluation of possible HO conditions, thus can trigger proactive measurement reports from the UE and/or proactive HO actions at the serving gNB.
These two steps are repeated with a time periodicity given, e.g., by the sampling rate and time filtering of the input RSRP measurements \cite{TS38.331_2021}, or alternatively, the steps can also be triggered by the serving gNB when certain traffic or mobility Quality-of-Service (QoS) conditions are met.

The outlined ML-based mobility algorithm can be implemented in either the UE or gNB or both, depending on the available ML assistance capabilities in each node. Furthermore, the mechanism can be integrated in self-organizing network-based Mobility Robustness Optimization solutions. 

\subsection{CSI Feedback}\label{CSI_Feedback}
CSI feedback in the downlink channel is a major challenge in Release\,$17$ and beyond. Currently, CSI precision is affected by compressing the measurements imposed by the standard. 

\begin{figure}
  \centering
  \subfigure[Trend of MSE.]{\includegraphics[scale=0.57]{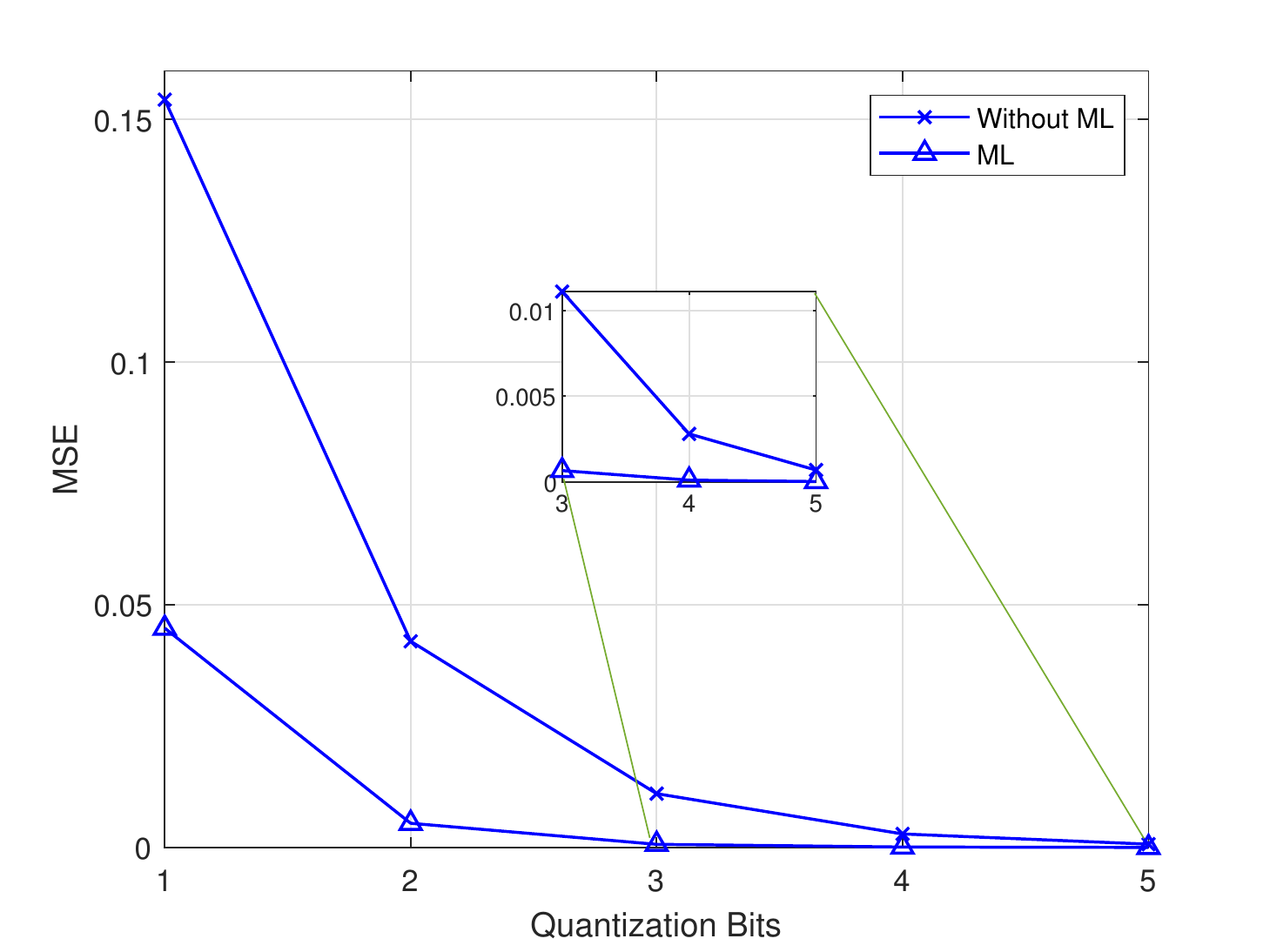}\label{fig:exph}}
  
  \subfigure[Trend of precoding gain.]{\includegraphics[scale=0.57]{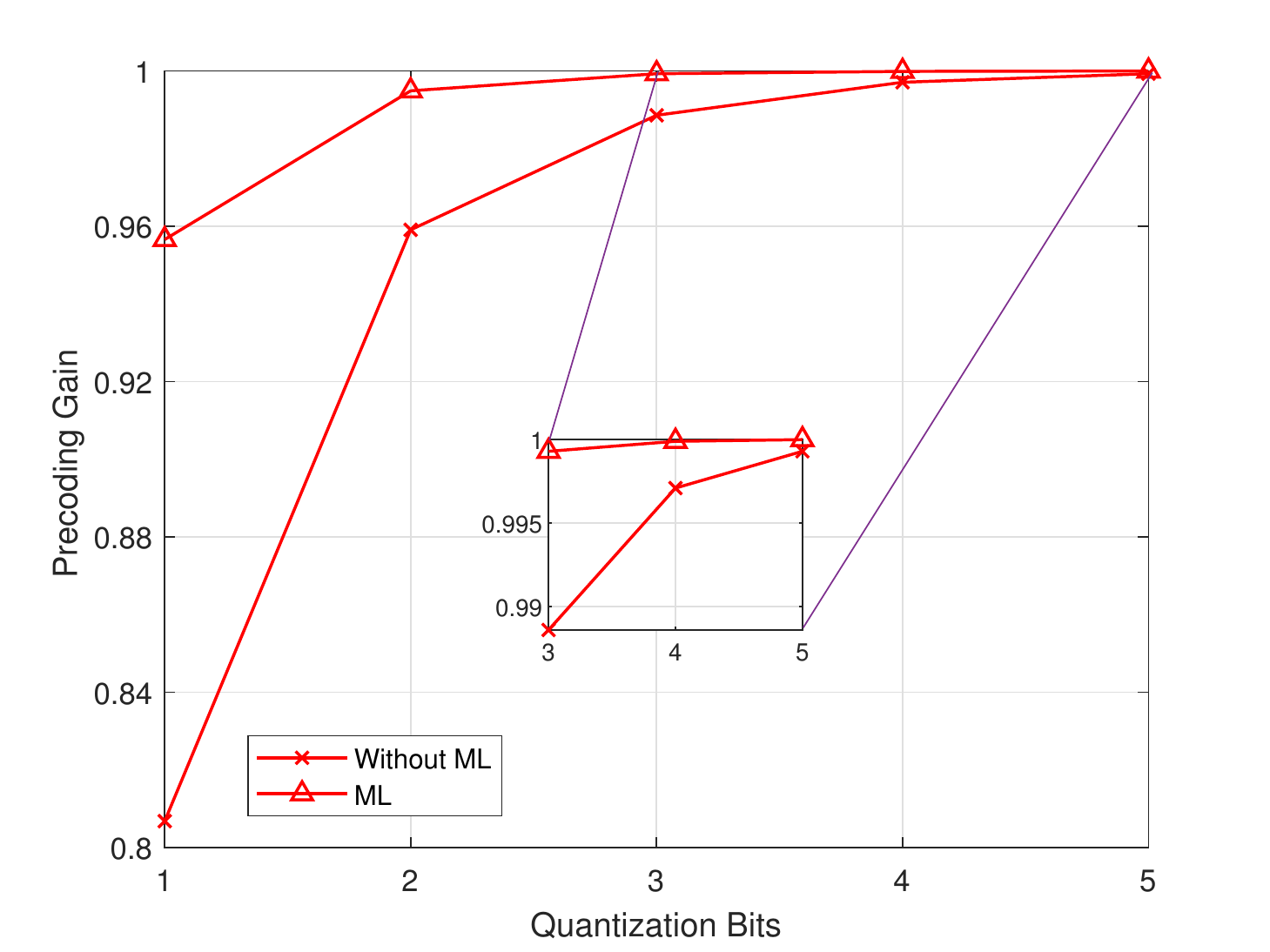}\label{fig:exgr}}
    
  \caption{Performance of MSE and precoding gain. $2\times1$ MIMO configuration is considered, and RNN is composed of $1$ hidden layer. For parameters’ details, refer to \cite{RNN}.}
  \label{CSI}
\end{figure}

In our study, summarized in Section\,\ref{MLPHY}, we assumed two RNN-based \textit{twin} channel predictors at the gNB and UE \cite{RNN}. The past CSI is utilized for training the RNN at both ends of the communication system. UE's feedback is evaluated with respect to the predicted channel. Fig.\,\ref{CSI} depicts the mean-squared error (MSE) between the actual channel versus the acquired channel at the gNB and the precoding gain when different quantization bits are used to feedback the CSI from the UE. The results are compared with and without using ML for the CSI feedback. A clear benefit of using ML can be observed. We believe that ML-based solutions will improve current performance without increasing signaling overhead.

\begin{figure*} 
\begin{center}
  \includegraphics[width=12cm,height=8cm]{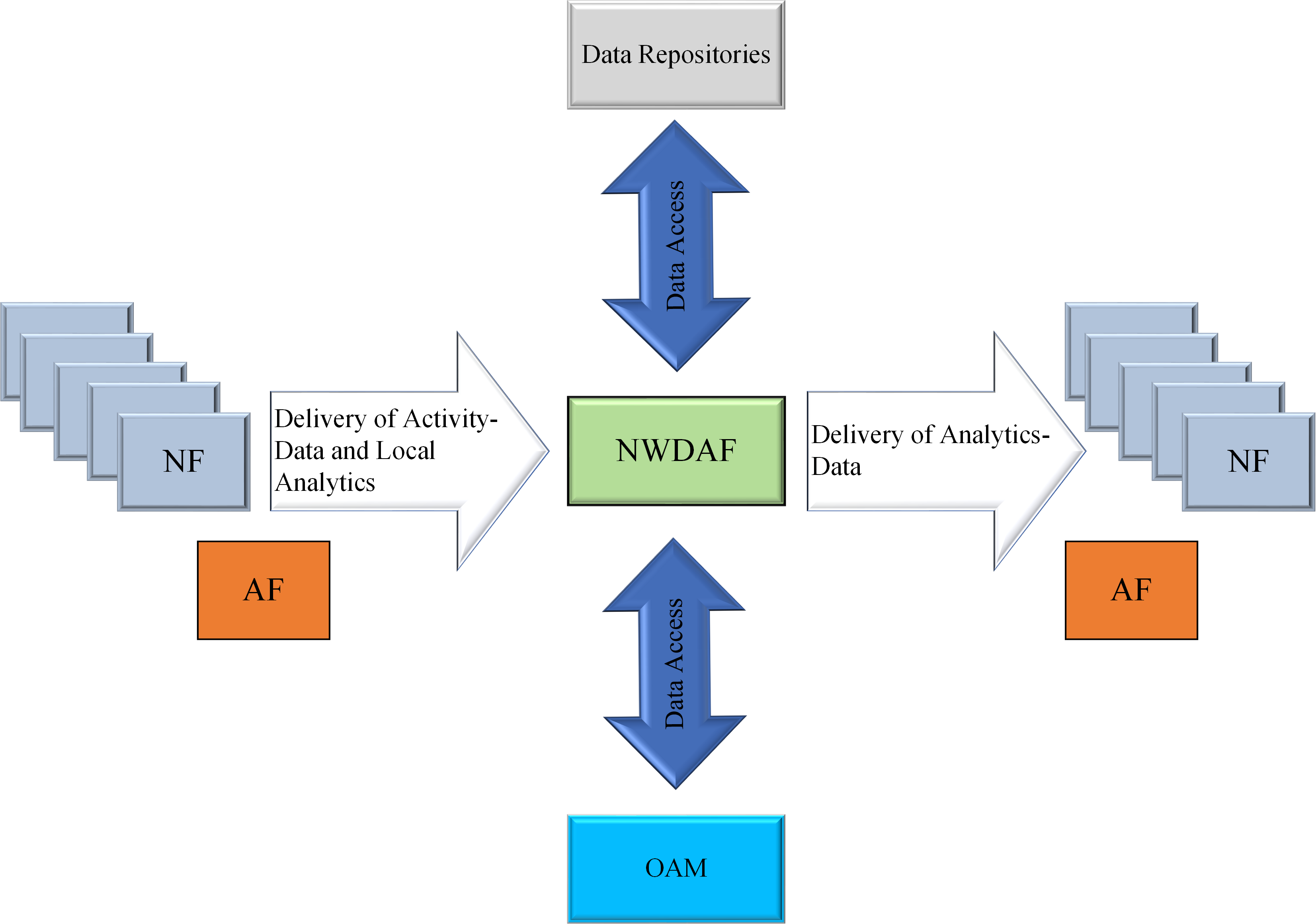}
  \caption{A generalized framework for 5G network automation in Release\,$16$, representing that NWDAF should be able to collect data from the operator OAM, AFs and 5GC network functions \cite{3GPPAI}.}\label{network_auto}      
\end{center}
\end{figure*} 

\section{Role of ML in Standardization }\label{sec4}

The potential of ML for 5G has been widely acknowledged in the literature and applications made it even in the standard at higher levels, e.g., for networking and security \cite{3GPPAI}. 3GPP has introduced a specification, named network data analytics function (NWDAF), in Release\,$15$ and $16$, as part of the 5G Core (5GC) architecture \cite{3GPPAI}. NWDAF is responsible for providing network analytics when requested by a network function (NF). Data is collected via application function (AF), operation, administration, and maintenance (OAM), NF, and data repositories. The specifications have also addressed the problem of inter-working for automation and data collection, which analytics vendors previously faced. 3GPP NWDAF framework for 5G systems is depicted in Fig.\,\ref{network_auto}. This automation gives leverage to network vendors for the deployment and testing of non-RT ML-related use cases. In Fig.\,\ref{network_auto}, inward interfaces aggregate data from different network sources, where communication occurs using existing service-based interfaces. Outward interfaces provide decisions (analytics-based, algorithmic) to AF and NF.

Regarding PHY, ML techniques lag behind, due to a number of issues. First, PHY makes use of abstractions and mathematical models that are inferred from the physical reality and electromagnetic principles. As long as such models describe the real-world precisely, there is no need for ML. Nevertheless, in practice, models and fixed algorithms are inefficient when facing rapidly changing and heterogeneous environments. For example, using the same channel acquisition scheme to acquire CSI from a laptop in line-of-sight with a gNB, a tablet on a fast train, or a mobile quickly moving in a super densely covered area might not be optimal. Consequently, the standardization efforts of intelligent techniques have gained momentum, and while 3GPP is ready to begin a study item on ML implementations, open-radio access network (O-RAN) will be ML-native, defining a RAN intelligent controller (RIC), which will enhance several RAN functions.

3GPP has started studying the implications of the ML use at layer-1 and a study item on ML for NR air interface has been agreed upon. After the RAN-1 working group studies, protocol aspects will be studied in RAN-2 and subsequently, interoperability and testability aspects will be considered in RAN-4 working group. The remaining part of this section summarizes the status of the standardization of ML techniques for PHY for both 3GPP and O-RAN.

\subsection{CSI Feedback}
CSI feedback for downlink channel in Release\,$17$ is a complex issue in which UE-based beam selection is followed by CSI reference symbols (RS) training and precoding matrix index (PMI) reporting, and lastly by Demodulation Reference Signal (DMRS) and consequent estimation of the precoded channel. Broadly, beam selection aims to establish a sufficiently strong link budget between the UEs and the gNB. The CSI-RS is used for \textit{fine} channel estimation, which is then fed back to the gNB to compute a precoder (eventually multi-user); finally, DMRS are precoded pilots that the UEs use to implement coherent demodulation. 
Currently, each of these phases is created following pre-established rules, with little to none room for intelligent behaviour. 
ML has been envisioned to possibly enhance each phase in a different way. Beam selection can be improved by intelligently correlating the beams with position or identity of the UEs. This would allow for a smart selection of the beams from the gNB side, thus avoiding brute-force selection. The CSI-RS can be enhanced by compressing the pilots and the PMI feedback exploiting \textit{ad hoc} ML compressors. Furthermore, channel prediction techniques \cite{CP_Karam} can be used in order to pre-establish a baseline for the CSI feedback \cite{RNN}.
Other aspects that can be improved include frequency of pilots in both CSI-RS and DMRS, power and timing and CSI-RS port selection. 

\subsection{RS-DMRS}
Roughly speaking, DMRS are RS used for channel estimation to perform coherent demodulation. The correct estimation of the channel using such pilots have a strong impact on the performance in terms of bit-error-rate and thus block-error-rate. The role of the ML in such domain is twofold. First, it can be used to improve the performance of the channel estimation.
Second, the ML can provide a smarter positioning of DMRS in order to reduce their frequency, hence reducing the overhead footprint in 6G. 

\begin{figure*} 
\begin{center}
  \includegraphics[width=12cm,height=7cm]{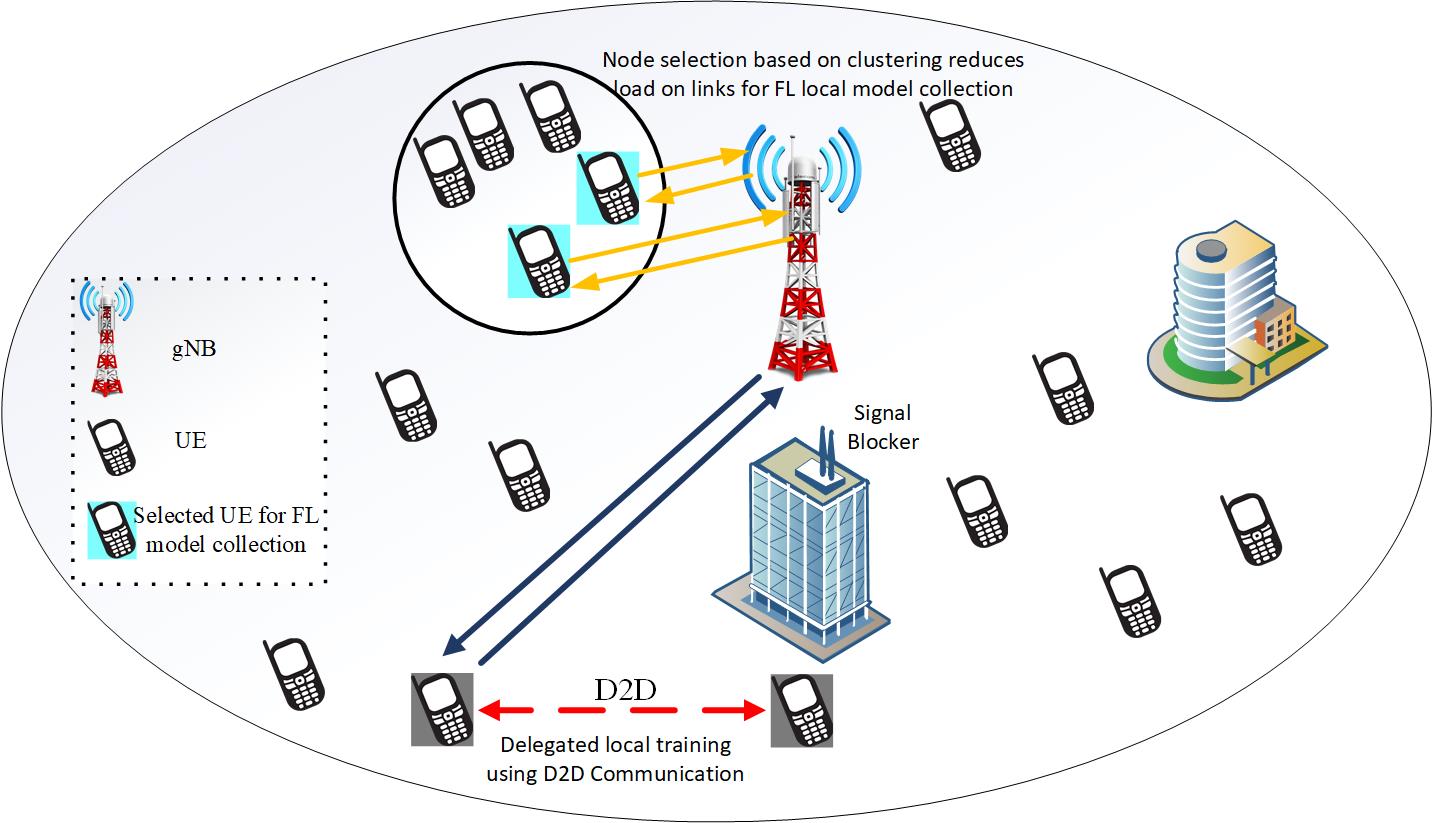}
  \caption{Model collection for FL in a wireless network when some of the UEs have large blockage and use D2D communication for model transfer. Cluster-based UE selection is another solution for asynchronous model collection to meet network QoS requirements.}
  \label{Majid_FL}.      
\end{center}
\end{figure*} 
\subsection{Positioning}
A precise positioning is one of the aspects that sees the largest improvement with respect to LTE's observed time difference of arrival (OTDOA) and uplink time difference of arrival (UTDOA), defined in Release\,$9$ onward. Various aspects of 6G allow for precise positioning of the UE, such as large number of antenna elements at the gNB, millimeter wave transmissions, dense network deployment. However, the methods based on angle-of-arrival and time-of-arrival fall short when non-line-of-sight scenarios are considered, in  interference-limited scenarios. 
ML techniques, see Fig.\,\ref{Fig2}, are expected to help in improving the position by exploiting channel charting, 
hence learning the likely position of a UE based on a report, and multiplexing together information that carries positioning information but are hard to exploit in a classical way, such as CSI report and sounding reference signal maps.

\subsection{Mobility Enhancements}
In 6G, 
frequent cell-selection, and frequent RSRP measurement could impact UEs' battery life. Furthermore, load balancing algorithms can use intelligent techniques that exploit the UE specific channel prediction, movement trajectory prediction and traffic demands prediction. 
Furthermore, the scenarios like fast-trains or non-terrestrial networks, will pose challenges to HO and conditional-HO operations. Novel solutions envisaged, compared to current 3GPP Release\,$17$, include the use of UE specific ML-based predictive algorithms, addressed in Section\,\ref{sec:mobility}, designed to reduce paging errors and HO failures; thus, improve the overall QoS.

\subsection{Standardization for ML Data Collection}
3GPP has started working on data collection for running ML algorithms in 5G networks \cite{3GPPAI3}. The scope of such studies include identifying mechanisms to collect data from the network through minimization of drive test framework or further advanced enhancements. Furthermore, studies will focus on discussing hosting of ML models both for training as well as inference purposes at various network entities for various use cases and defining any new interfaces required for transporting data to the models.

\subsection{Federated Learning Model Collection}
Training and prediction based on ML models will put an extra load on networks already transporting a large volume of data. Therefore, it is important to estimate the effect of model training and inference on network traffic, particularly for federated learning (FL) where UEs will act as distributed hosts \cite{3GPPAI2}. The latency in collecting locally trained models is bounded in FL and network links should be able to meet delay budgets. This is particularly challenging in today's networks where a UE's own QoS requirements are already demanding and the FL model training and collection will further incur an extra burden on the network. Similarly, the split inference, where UEs cooperate with each other to perform joint inference, results in increasing the network traffic.
3GPP studies in Release\,$18$ \cite{3GPPAI2} will focus on the above mentioned issues to support training and inference for ML/FL models over wireless links.   

\subsection{O-RAN-RIC}
O-RAN alliance, aims to define a RAN network that is non-vendor specific, and that has an innate support for ML as an enabler for automation and OPEX savings. O-RAN alliance has defined interfaces for exchange of information in the protocol stack. To this end, in the O-RAN architecture, ML-assisted RAN intelligent controller (RIC) is included for network automation, for both scenarios, i.e., non-RT and RT. In the non-RT RIC, ML algorithms' training is done by using the data obtained at lower layers. However, the learning process remains slow; therefore, it is called non-RT RIC. Later, the learner is fed into the RT RIC, which utilizes the RT captured data to perform decisions online. Additionally, the functionality of non-RT includes policy management and higher layer procedure optimization. Therefore, the RAN or core-network can deploy such a mechanism based on the collected data.

\section{Open Challenges and Roadmap for Deploying ML Techniques} \label{sec5}
Though ML is a potential technology and enabler for next-generation wireless networks, several challenges related to its practical use are addressed below.
       
\subsection{Data Availability and Benchmarking}
One of the foremost challenges in wireless networks is \textit{data availability}. Data availability concerns the problem of identifying a common and accepted set of data (e.g., channel realizations) with the goal of testing and benchmarking 
ML algorithms. Such a problem is of a pivotal importance for standardization, where normally algorithms and proposals are tested using agreed underlying physical models (e.g., urban macrocells/microcells channel models), evaluation methodologies and calibrated simulators. Contrary to other fields, 
cellular networks have no standard data set to train and benchmark an ML algorithm. Therefore, a synthetic data set or software generated data set
is of a predominant importance to train and benchmark ML algorithm(s), and to agree on a common evaluation methodology to rank proposition and standard algorithms.

Identifying a set of key performance indicators in wireless networks is another crucial task for ML standardization. It is necessary to design a set of metrics to classify and rank ML algorithms and their performance. Classic approaches such as throughput and signal-to-interference-plus-noise ratio (SINR) might not be sufficient since a small improvement in these values might come at the cost of large complexity augmentation and exacerbated energy consumption. 

\subsection{Selection of ML versus Non-ML Solutions}
ML tools are regarded as an implementation-oriented tool rather than a standard relevant aspect. The idea behind this relies on the fact that each vendor has the freedom to efficiently implement each aspect of the standard as long as the external interfaces are respected. A simple example of this is given in the CSI feedback, where a UE needs to select a specific PMI, but the standard does not specify any specific way in which this selection is performed. Recently, however, the idea of having ML dedicated message exchanges and performance that only an ML-aided algorithm can achieve has paved the way for standardization of ML algorithms \cite{RNN}. This opens the door for several issues, e.g., will the standard impose a specific ML structure, classifying minimum performance and implementation structure, or will it remain far from the implementation? With regards to NNs, it is still open if hyperparameters are going to be left to vendor-specific implementation or will they be set by the standard.
\subsection{Complexity of ML Algorithms}
Considering the limited battery life, storage, computational capability, and limited communication bandwidth in most cellular network entities, an ML model's cost-performance tradeoff becomes a fundamental issue. Another issue is the speed/time-steps at which the training and inference needs to be performed. Whereas hard-wired gNB have sufficient computational power to run complex ML algorithms, UEs need to face battery, heating and stringent complexity limits. 
Possible solutions to such issue include, but not limited to implementation of substitute rule-based algorithms at the UE side, migrating the load all on the gNB side.
\subsection{Communication-aware Federated Learning}\label{Federated_Learning}
Traditional ML models support centralized learning. Due to difficulties in collecting large amount of training data from the UEs, privacy issues and bandwidth bottleneck, FL has emerged as a promising solution.
In FL, training is performed distributively over network devices, called local model hosts, and an application server on the network side acts as a central host to aggregate local models transmitted by the local learners. 
Typically, an application server host aggregates models only when updates are available from all the local learners, called synchronous model transfer. However, this is highly inefficient in wireless networks where links are unpredictable, local learners (UEs) are energy limited and have their own QoS requirements. Asynchronous model collection is the most viable solution for FL in wireless networks, where a subset of UEs is selected for a local model update in each round of model collection. However, UE selection in each round is a complex problem because UEs are energy limited and the network bandwidth is scarce, hindering collection of local models from all the UEs to represent independently and identically data collection. These mechanisms are usually vendor proprietary, but standardization still needs to define some common mechanisms for efficient model collection. As shown in Fig.~\ref{Majid_FL}, UE clustering and local device-to-device (D2D) communication for asynchronous model collection are possible solutions to decrease network communication and will require standardization support. 
\subsection{Stability and Adaptability of ML Techniques}
ML algorithms applied to wireless networks must be adaptive as they will have to deal with parameters that change dynamically. 
Particularly, the weights of the NN are evaluated online based on the trained data. However, this approach may not be applicable in wireless, and specifically in a standard, where coordination among entities belonging to different operators and provided by different vendors have to coexist, and in which the need for quick response could prevent one or the other solution. Possible solutions include: pre-trained NN, or partially trained NN (i.e., NN in which the starting point is pre-set); cloud-based downloadable data set for NN training; codebook-based NN, in which a codebook of different NNs is used and agreed upon between the gNB and UEs.
Another related problem is to detect an outdated ML model with high inference error and replace it. Replacing an outdated model with a new model incurs further delay.
Thus, there must be a proactive mechanism to adapt the ML model to network conditions such that network functions suffer minimum performance loss.

\section{Conclusion}\label{conclusion}
Motivated by the promise of the use of ML algorithms, we presented an overview of ML techniques to be used in 5G-Advanced and 6G wireless networks. Furthermore, we discussed the key roles of ML-based solutions from industrial and standardization perspectives. We also highlighted the practical challenges of deploying ML techniques in wireless networks and how to deal with them. 
Non-RT and higher layer ML-based solutions can be, and are, applied already in today's networks. Implementing RT ML solutions at PHY/MAC in 6G networks are the next big challenge in the research community.
We believe that overcoming these challenges, both in research as well as at standardization levels, will pave the way for next-generation wireless communication to be effective and sustainable. 
\ifCLASSOPTIONcaptionsoff
  \newpage
\fi
\bibliographystyle{IEEEtran}
\bibliography{References}
\vspace{-0.5cm}
\begin{IEEEbiography}[{\includegraphics[width=1in,height=1.25in,clip,keepaspectratio]{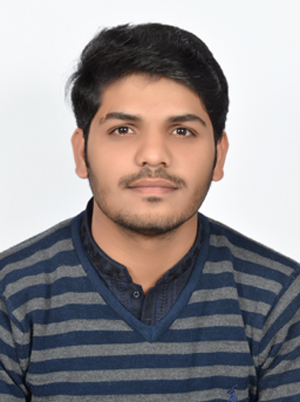}}]{Muhammad K. Shehzad} 
[S'21] is working as a Research Engineer and Ph.D. student at Nokia Bell-Labs and CentraleSupelec, Paris, France, respectively. He received his B.Eng. (Hons.) degree  in Electrical and Electronic Engineering from the University of Bradford, Bradford, U.K., in $2016$, and M.S. in Electrical Engineering from the National University of Sciences \& Technology (NUST), Islamabad, Pakistan, in $2019$. His major research interest is in MIMO communication using Artificial Intelligence (AI)/Machine Learning (ML).
\end{IEEEbiography}
\vspace{-0.5cm}
\begin{IEEEbiography}[{\includegraphics[width=1in,height=1.25in,clip,keepaspectratio]{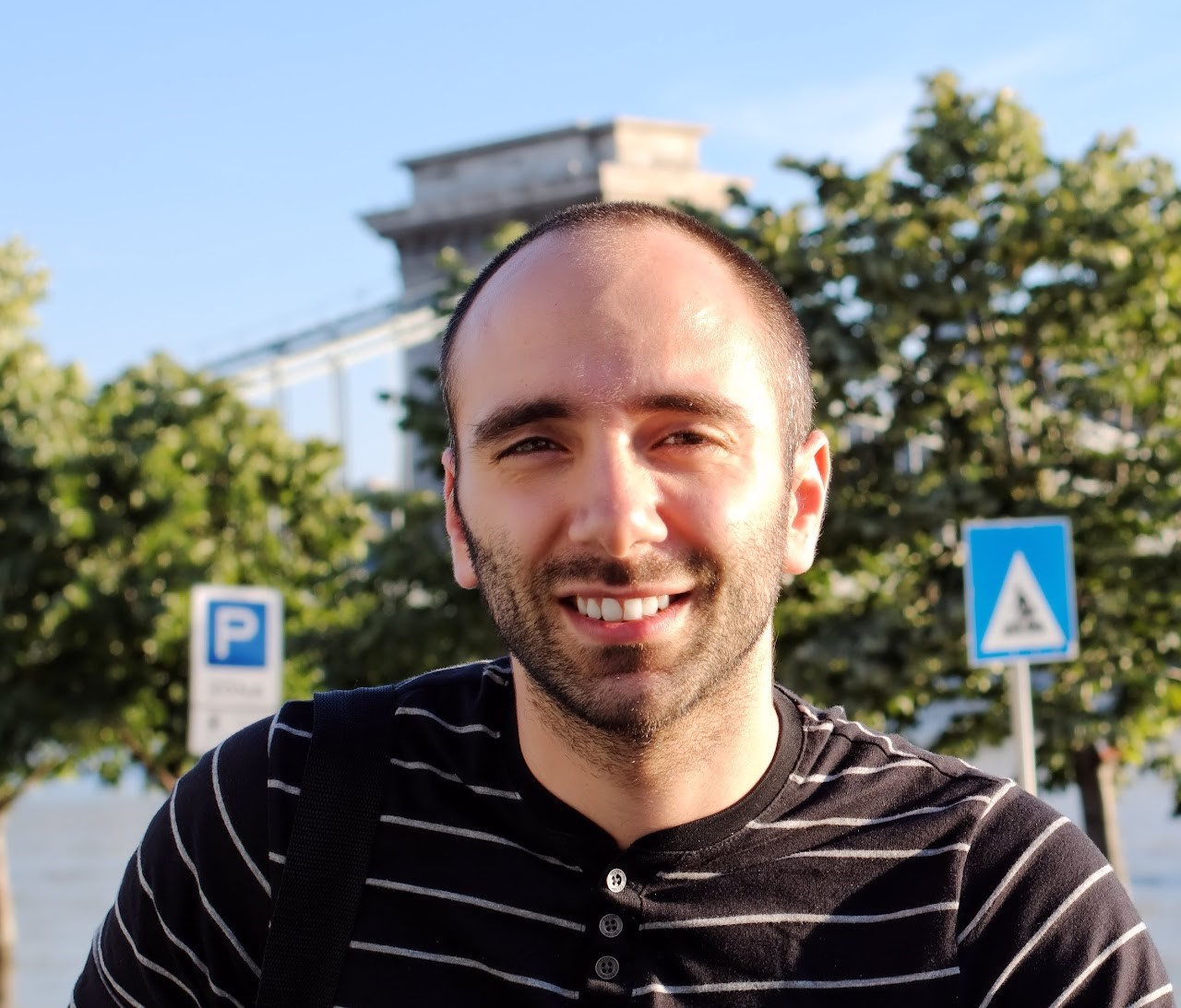}}]{Luca Rose}
[M'11] is Senior research and standardization expert with Nokia Bell-labs. He received his M.Sc. from university of Pisa, Italy, and his Ph.D. in Physics from Centrale-Supelec. He worked with Huawei France research center and Thales Communications and Security, contributing to several standard organizations. He is currently an ITU-R and ETSI delegate  and the lead editor of IEEE Communication magazine series on IoT. His interests span from the field of AI/ML to Game theory. \end{IEEEbiography}
\vspace{-0.5cm}
\begin{IEEEbiography}[{\includegraphics[width=1in,height=1.25in,clip,keepaspectratio]{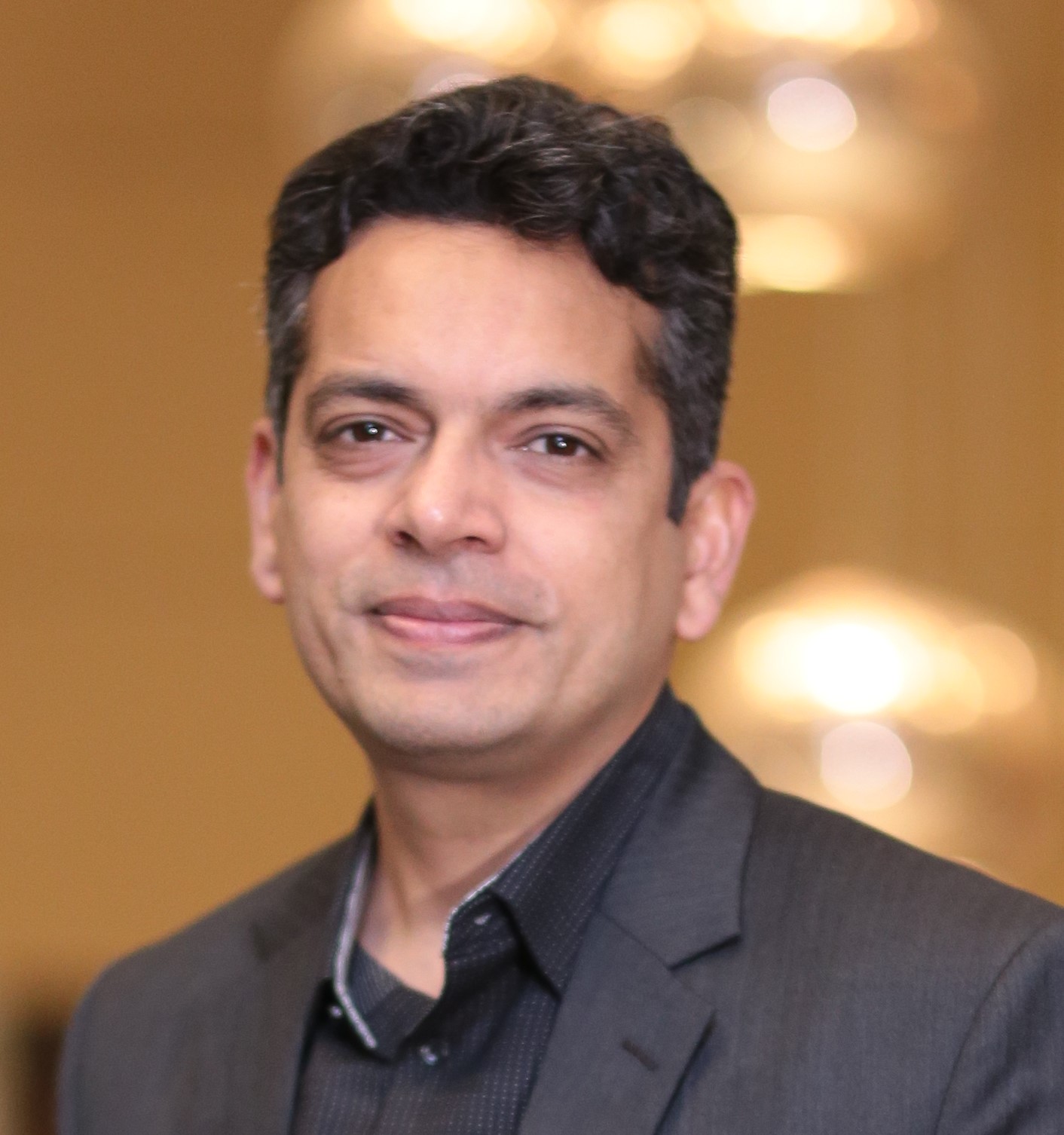}}]{M. Majid Butt}
[SM'15] is a Senior Specialist at Nokia Bell-Labs, France, and an adjunct Professor at Trinity College Dublin, Ireland. He has authored more than $70$ peer-reviewed conference and journal articles and filed over $30$ patents. He is IEEE Comsoc distinguished lecturer for the class $2022$-$23$. He frequently gives invited and technical tutorial talks on various topics in IEEE conferences and serves as an associate editor for IEEE Communication Magazine, IEEE Open Journal of the Communication Society and IEEE Open Journal of Vehicular Technology.
\end{IEEEbiography}
\vspace{-0.5cm}
\begin{IEEEbiography}[{\includegraphics[width=1in,height=1.25in,clip,keepaspectratio]{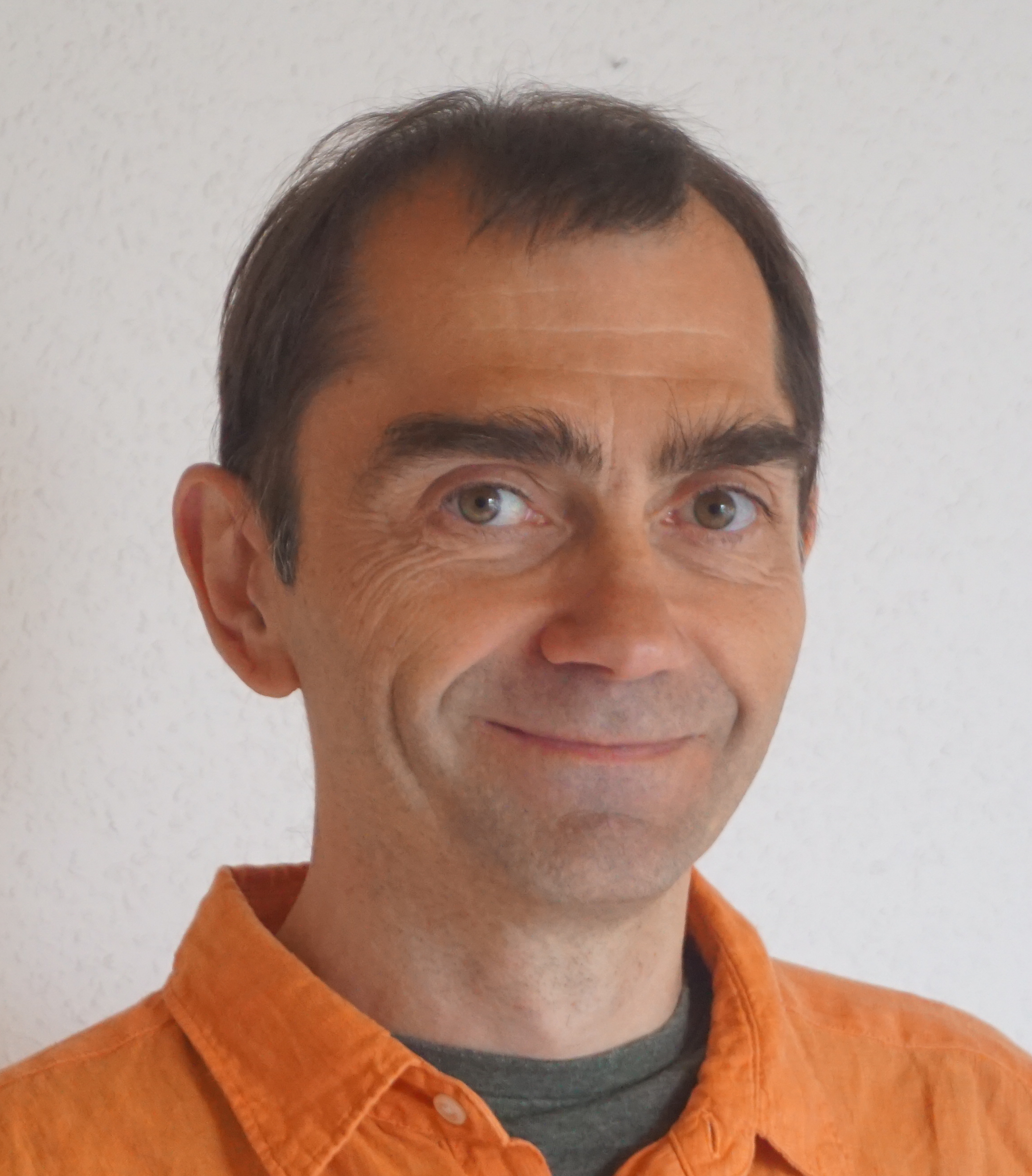}}]{István Z. Kovács}
[M'00] received his B.Sc. from “Politehnica” Technical University of Timişoara, Romania in 1989, his M.Sc.E.E. from École Nationale Supérieure des Télécommunications de Bretagne,  France in $1996$, and his Ph.D.E.E. in Wireless Communications  from Aalborg University, Denmark in 2002. Currently he is senior research engineer at Nokia, Aalborg, Denmark, where he conducts research on machine learning-driven radio resource management and radio connectivity enhancements for non-terrestrial and aerial vehicle communications, in LTE and 5G networks.
\end{IEEEbiography}
\vspace*{5mm}
\begin{IEEEbiography}[{\includegraphics[width=1in,height=1.25in,clip,keepaspectratio]{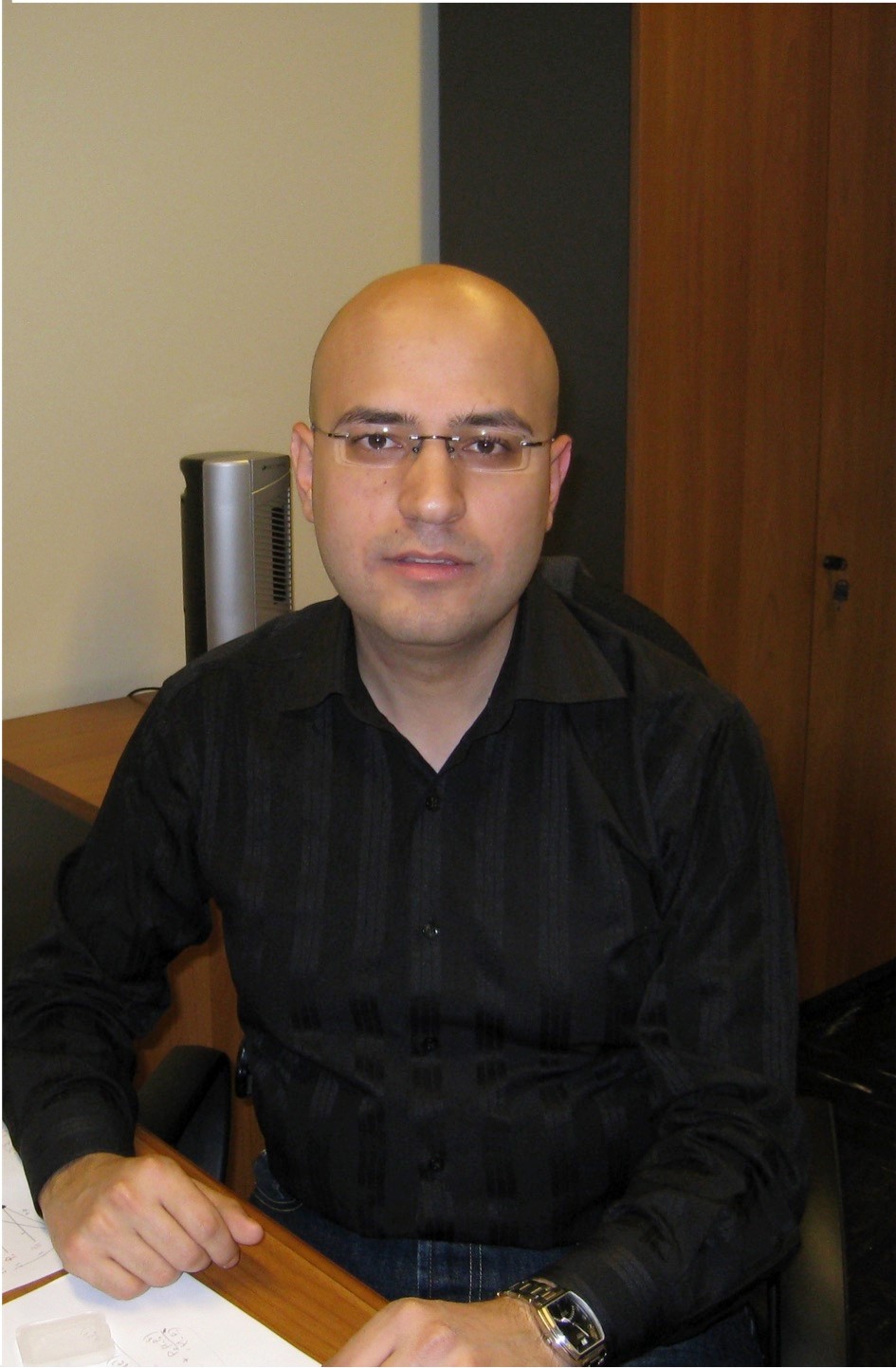}}]{Mohamad Assaad}
[SM'15] is a Professor at CentraleSupelec, France and a researcher at the Laboratory of Signals and Systems (CNRS). He has co-authored $1$ book and more than $120$ journal and conference papers and serves regularly as TPC co-chair for top-tier international conferences. He is currently an Editor for the IEEE Wireless Communications Letters and Journal of Communications and Information Networks. His research interests include 5G and beyond systems, and Machine Learning in wireless networks.
\vspace{0mm}
\end{IEEEbiography}

\begin{IEEEbiography}[{\includegraphics[width=1in,height=1.25in,clip,keepaspectratio]{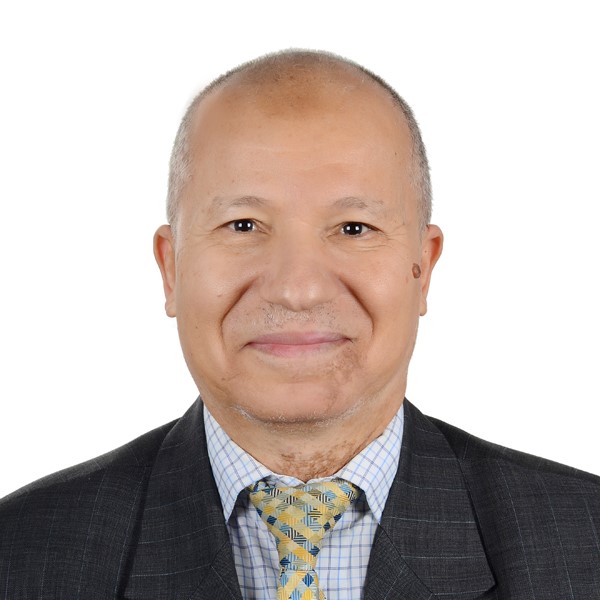}}]{Mohsen Guizani}
[F’09] is currently a Professor at the Machine Learning Department at the Mohamed Bin Zayed University of Artificial Intelligence (MBZUAI), Abu Dhabi, UAE. His main research interests are wireless communications and IoT security. He was elevated to the IEEE Fellow in $2009$. He was listed as a Clarivate Analytics Highly Cited Researcher in Computer Science in $2019$, $2020$ and $2021$. Dr. Guizani has won several research awards. He is the author of ten books and more than $800$ publications.
\end{IEEEbiography}
\end{document}